\documentclass[%
 aip,
 jmp,%
 amsmath,amssymb,
nofootinbib,
preprint,%
]{revtex4-1}

\usepackage{graphicx}% Include figure files
\usepackage{dcolumn}% Align table columns on decimal point
\usepackage{bm}% bold math
\def\VEC#1{\mbox{\boldmath $#1$}}

\begin{document}

\preprint{}

\title[Charge separation instability around black hole]
{Charge separation instability in an unmagnetized disk plasma \\
around a Kerr black hole} % Force line breaks with \\

\author{Shinji Koide}
\affiliation{Department of Physics, Kumamoto University, 
2-39-1, Kurokami, Kumamoto, 860-8555, JAPAN}
\email{koidesin@sci.kumamoto-u.ac.jp} %Lines break automatically or can be forced with \\

\date{\today}% It is always \today, today,

\begin{abstract}
In almost all of plasma theories for astrophysical objects, we have assumed 
the charge quasi-neutrality of unmagnetized plasmas in global scales. 
This assumption has been justified
because if there is a charged plasma, it induces electric field 
which attracts the opposite charge, and 
this opposite charge reduces the charge separation. 
%% This quasi-neutrality concept can be generalized for strongly magnetized 
%% relativistic plasmas, which are charged stationarily.
Here, we report a newly discovered instability which causes
a charge separation in a rotating plasma inside of an innermost stable
circular orbit (ISCO) around a black hole.
The growth rate of the instability is smaller than that of the disk instability 
even in the unstable disk region and is forbidden in the stable disk
region outside of the ISCO. 
However, this growth rate becomes comparable to
that of the disk instability when the plasma density is much lower than 
a critical density inside of the ISCO.
In such case, the charge separation instability would become apparent and
cause the charged accretion into the black hole, thus charge the hole.
\end{abstract}

%Use showkeys class option if keyword
\maketitle

% \begin{quotation}
% The ``lead paragraph'' is encapsulated with the \LaTeX\ 
% \verb+quotation+ environment and is formatted as a single paragraph before the first section heading. 
% (The \verb+quotation+ environment reverts to its usual meaning after the first sectioning command.) 
% Note that numbered references are allowed in the lead paragraph.
% %
% The lead paragraph will only be found in an article being prepared for the journal \textit{Chaos}.
% \end{quotation}

\section{Introduction} \label{sec1}

In a scale larger than several Debye lengths, unmagnetized plasmas in the Universe 
have been assumed to be quasi-neutral in charge. 
This is because even a small charge imbalance would result in very large electric fields 
which would 
%% act to move the plasma charged particles so as to
cause the very small time scale electrostatic oscillation called plasma oscillation,
and the damping or averaging of the oscillation 
over the mesoscopic time scale restores neutrality 
%% very quickly 
\citep[e.g.,][]{bellan06,frank85}.
On the other hand, if we assume the ideal MHD condition on a magnetized, relativistically 
moving plasma, the charge density becomes significant, and this charge density
is called the Goldreich-Julian density, $\rho_{\rm GJ}$ \citep{goldreich69}.
It is noted the Goldreich-Julian density is required so that the electric field induced 
by the charge density vanishes in the comoving frame of the plasma.
%% if the particle number density of the plasma is large enough.
In such case, it seems that we cannot assume the charge quasi-neutrality.
However, we point out here that the difference between the net charge density
of the magnetized plasma $\rho_{\rm e}$ 
and the Goldreich-Julian density, 
$\Delta \rho_{\rm e} = \rho_{\rm e} - \rho_{\rm GJ}$, plays the same role
as the charge in the unmagnetized plasma. 
In this paper, we call the difference $\Delta \rho_{\rm e}$ the ``free charge density",
which induces the electric field observed by the comoving frame of the plasma.
In the scale larger than several Debye lengths, the  free charge density
should tend to vanish
because of the same reason for the charge quasi-neutrality of the unmagnetized plasma.
This can be regarded as the generalization of the concept of charge quasi-neutrality
to the magnetized, relativistic plasma.
We call this concept ``free charge quasi-neutrality".
When the magnetic field is so strong and the thin plasma rotates so fast
as assumed in a pulsar magnetosphere that the number density of the 
plasma particles is 
smaller than the Goldreich-Julian density divided by the elementary electric charge $e$, 
the free charge density becomes significant and
the electric field observed by the comoving frame of the plasma remains.
The parallel component of this electric field to the magnetic field 
accelerates plasma particles directly.
%% because the electric field component parallel to the magnetic field becomes significant.
In the region of the remaining one-direction electric field component along the magnetic field, 
plasma is swept by the electric field 
and the vacuum called ``outer gap" appears \citep{holloway73,holloway81}.
Even in such vacuum, when plasma enters into it, the plasma is 
separated into positively and negatively
charged fluids and the two fluids move to the opposite directions along the magnetic 
field to decrease the electric field component.
Thus, in a strongly magnetized, relativistic plasma, the free charge tends to be
canceled to keep the free charge neutrality.
%% The outer gap contains no plasma unless additional plasmas enter into the region.
%% Then, even in the pulsar magnetosphere, plasma have no electric field
%% which accelerates the plasma particles stationarily so as to restore the active
%% charge quasi-neutrality.

%% It is summarized emphatically that in a scale larger than several Debye lengths,
%% the plasma has been assumed to satisfy the (active) charge quasi-neutrality
%% because plasma charged components move so that the electric field 
%% accelerating the plasma charged components vanishes on the average.
According to the above consideration, in a scale larger than several Debye lengths,
it has been assumed that the charged components of plasma move so that the electric field
accelerating the charged components decreases so as to restore the free
charge quasi-neutrality.
%% This is true except for plasmas rotating near black holes, as mentioned below.
%
Here, we report a charge separation instability in an unmagnetized plasma 
rotating around a black hole, which will induce the free charge density and 
electric field exponentially.
To investigate the charge separation of plasmas around the
black holes, we use generalized GRMHD equations derived by \citet{koide10}.
We present the linear analysis of the charge separation of plasmas
near Kerr black holes. We found the well-known plasma oscillation 
in the stable disk region outside of an innermost stable circular orbit (ISCO) 
around a black hole.
On the other hand, in a circularly rotating plasma inside of the ISCO, 
we found an instability of the charge separation.
The charge separation instability does not happen in the stable disk region
outside of the ISCO. Furthermore, even in the unstable disk inside
of the ISCO, the growth rate of the charge separation is smaller 
than that of the disk instability. However, when the plasma density is 
much lower than the critical density (the very low plasma density case),
the growth rates of the two instabilities become comparable and the
charge separation instability becomes apparent.
That is, due to the charge separation instability, the unstable
disk falling into the black hole can be charged.
%% The growth rate of the instability of the charge separation is smaller than
%% that of the disk instability inside of the ISCO. However, when the plasma density
%% is much lower than a critical density, the both growth rates of the charge
%% separation instability and disk instability become comparable and the charge separation
%% becomes significant.
%% This effect would cause the charged plasma falling into the black hole 
%% inside of the ISCO.

%% The Ohm's law in the generalized GRMHD equations indicates electromotive forces 
%% due to gravitation, centrifugal force, and 
%% frame-dragging effect around a black hole. 
%% The gravitational electromotive force was also pointed out by \citet{khanna98}.
%% The gravitational electromotive force can cause the magnetic reconnection
%% even in a case of zero resistivity (see the detail in \citet{koide10}).
%% A charge separation in a strong gravity is required to induce
%% the gravitational electromotive force.
%% The charge separation instability can supply the significant charge density
%% which will result in the magnetic reconnection.
%% The distinct mechanism is discussed in Section \ref{sec3}.
%% %% However, the charge separation is usually canceled by the oppositely
%% %% charged plasma attracted by the electric field due to the charge separation, 
%% %% and it causes the plasma oscillation \citep{bellan06}.
%% %% Then, such gravitational magnetic reconnection may not be stationarily caused 
%% %% over the plasma oscillation time. 

In Section \ref{sec2}, we present a linear analysis of charge separation in
a stationarily rotating disk plasma around a Kerr black hole
with a brief summary of the generalized GRMHD equations.
The analysis shows an instability of charge separation in the plasma inside of
the ISCO around a black hole.
In Section \ref{sec3}, we summarize the results and discuss briefly 
the astrophysical meanings of the charge separation instability.

\section{Charge Separation in Plasma Disk around Black Hole \label{sec2}}

We investigate the simplest process of charge separation in an unmagnetized disk rotating 
around a Kerr black hole. 

\subsection{Brief summary of generalized GRMHD equations}

In this subsection, we briefly summarize the generalized GRMHD equations;
see \citet{koide10} in more detail.
We investigate a charge separation using the generalized GRMHD equations
of plasmas in the space-time, $x^\mu =(t,x^1,x^2,x^3)$
around a black hole where a line element $ds$ is given by 
$ds^2 = g_{\mu \nu} dx^\mu dx^\nu$
(Equations (18), (24), and (59) with Equations (25) and (58) of \citet{koide10}).
Throughout this paper, except for a paragraph in Section \ref{sec3}, 
we use the unit system where light speed is unity and the energy densities
of electric field $\VEC{E}$ and magnetic field $\VEC{B}$ are given by 
$E^2/2$ and $B^2/2$ in the Minkowski space-time, respectively.

First of all, we summarize the generalized GRMHD equations for a general case 
briefly as follows:
\begin{eqnarray}
\nabla_\nu (\rho U^\nu) &=& 0 , \label{onefluidnum} \\
\nabla_\nu T^{\mu\nu} &=& 0 , \label{onefluidmom}  \\
 \frac{1}{ne}  \nabla_\nu \left ( \frac{\mu h^\dagger}{ne} 
Q^{\mu \nu} \right )
&=& U^\nu  {F^\mu}_\nu 
- \eta \left ( J^\mu - \rho_{\rm e}'  U^\mu \right )
+\frac{1}{2ne} \nabla^\mu (\Delta \mu p - \Delta p) \nonumber \\
&& - \frac{\Delta \mu}{ne} J^\nu {F^\mu}_\nu 
+ \eta \rho_{\rm e}' \Theta U^\mu ,
\label{onefluidohm}
\end{eqnarray}
and Maxwell equations
\begin{eqnarray}
\nabla_\nu \hspace{0.3em} ^*F^{\mu\nu} &=& 0 , \label{4formfar} \\ 
\nabla_\nu F^{\mu\nu} & = & J^\mu , \label{4formamp} 
\end{eqnarray}
where 
the energy-momentum tensor $T^{\mu\nu}$ and ``charge-current density tensor"
$Q^{\mu\nu}$ are given by
\begin{eqnarray} 
T^{\mu\nu} &\equiv& p g^{\mu\nu} + h^\dagger U^\mu U^\nu 
+ \frac{\mu h^\ddagger}{(ne)^2} J^\mu J^\nu 
+ \frac{2 \mu \Delta h^\dagger}{ne} (U^\mu J^\nu + J^\mu U^\nu ) 
\nonumber \\
&& + {F^\mu}_\sigma F^{\nu\sigma} - \frac{1}{4} g^{\mu\nu} 
F^{\kappa\lambda} F_{\kappa\lambda} , \\
Q^{\mu \nu} &\equiv& \frac{en}{\mu h^\dagger} K^{\mu \nu} 
= \frac{en}{h^\dagger} 
\left [ \frac{h^\ddagger}{ne} (U^\mu J^\nu + J^\mu U^\nu ) 
+ 2 \Delta h^\dagger U^\mu U^\nu
- \frac{\Delta h^\sharp}{(ne)^2} J^\mu J ^\nu \right ]   .  
\end{eqnarray}
Equation (\ref{onefluidohm}) presents the general relativistic 
generalized Ohm's law.
In Equation (\ref{onefluidohm}), the left-hand side expresses the inertia 
effect and transport of kinetic energy and momentum of the current,
the first two terms of the right-hand side
correspond to all terms of the ``standard'' Ohm's law with 
resistivity $\eta$, the third term
represents the thermo-electromotive force, the forth term expresses
the Hall effect, and the last term comes from the equipartition 
of the thermalized energy due to the friction force between
the two fluids. 
%% The variable $\eta$ indicates resistivity, 
Here, $\rho_{\rm e}'=-J^\nu J_\nu$ is the charge density observed 
by the local rest frame of the plasma and
$\Theta$ is the rate of equipartition
with respect to the thermalized energy due to friction 
(for detail, see Appendix A of \citet{koide09}).
We follow the notations used by \citet{koide10} with respect to 
physical variables except that we use $Q^{\mu \nu}$ instead of $K^{\mu \nu}$.
Here, we used the two-fluid model, where we assumed the plasma consists 
of positively charged particles with
charge $e$ and mass $m_+$ and negatively charged particles with charge $-e$
and mass $m_-$ (Appendix \ref{appenda}). 
We used the typical mass of a plasma particle $m \equiv m_+ + m_-$,
normalized reduced mass $\mu  \equiv  m_+ m_-/m^2$, and normalized mass difference
$\Delta \mu  \equiv  (m_+ - m_-)/m$. 
The variables $\rho$, $h^\dagger$, $p$, $n \equiv \rho/m$,
$\Delta p$, and $\Delta h^\dagger$ are the mass density, 
enthalpy density, pressure,
number density, pressure difference of two fluids, and 
difference of two fluid enthalpy density. Furthermore, $\nabla_\mu$,
$U^\mu$, and $J^\mu$ are the covariant derivative, 4--velocity, 
and 4--current densities, respectively, and $F_{\mu\nu}$ is the electromagnetic
strength tensor and $^*F_{\mu\nu}$ is the dual tensor of $F_{\mu\nu}$. 
Here, the electric field is given by $E_i = F_{i0}$ and
the magnetic field is $F_{ij} = \sum_k \epsilon_{ijk} B_k$
($\epsilon_{ijk}$ is the Levi-Civita symbol), 
where the alphabetic index ($i$, $j$, $k$) runs from 1 to 3. 
We also use the variables related to the enthalpy density,
\begin{equation}
h^\ddagger \equiv h^\dagger - \Delta \mu \Delta h^\dagger, \verb!    !
\Delta h^\sharp \equiv \Delta \mu h^\dagger 
- \frac{1-3\mu}{2\mu} \Delta h^\dagger.
\end{equation}
It is noted that Equation (\ref{4formamp}) yields the equation of continuity 
with respect to the current,
\begin{equation}
\nabla_\nu J^\nu = 0.
\label{concur}
\end{equation}

%%%% In the generalized equations and the Maxwell equations, there are three
%%%% types of terms including covariant derivatives: with respect to 
%%%% (i) contravariant vector like $\rho U^\mu$ and $J^\mu$, 
%%%% (ii) anti-symmetric 2nd rank tensor like $F_{\mu\nu}$ and $^*F_{\mu\nu}$,
%%%% and (iii) symmetric 2nd rank tensor like $T^{\mu\nu}$ and $K^{\mu\nu}$.
%%It is noted that as for any contravariant vector $A^\mu$ and 
%% any anti-symmetric
%%2nd rank tensor $A^{\mu\nu}$, we have
%%\begin{eqnarray}
%%\nabla_\nu A^\nu &=& \frac{1}{\sqrt{-g}} \partial_\nu \left (
%%\sqrt{-g} A^\nu \right ) , \label{eqcov} \\
%%\nabla_\nu A^{\mu\nu} &=& \frac{1}{\sqrt{-g}} \partial_\nu \left (
%%\sqrt{-g} A^{\mu\nu} \right ) , \label{eqast} 
%%\end{eqnarray}
%%where $g$ is the determinant of the metric $g_{\mu\nu}$.
%%With respect to an arbitrary symmetric 2nd rank tensor $S^{\mu\nu}$, 
%%the derivative is written by the Christoffel symbols,
%%$\Gamma^\lambda_{\mu\nu} = \frac{1}{2} g^{\lambda \sigma}
%%(-\partial_\sigma g_{\mu\nu} + \partial_\mu g_{\nu\sigma} 
%%+ \partial_\nu g_{\sigma \mu})$ as
%%\begin{equation}
%%\nabla_\nu S^{\mu\nu} = \frac{1}{\sqrt{-g}} \partial_\nu 
%%\left (
%%\sqrt{-g} S^{\mu\nu} \right ) + \Gamma^\mu_{\sigma\nu} S^{\sigma\nu}  .
%%\end{equation}

We assume that off-diagonal spatial elements of the metric $g_{\mu\nu}$
vanish: $g_{ij} = 0 $ $(i \ne j)$. Writing the non-zero components by 
$ g_{00}=-h_0^2$, $g_{ii}=h_i^2$, $g_{i0}=g_{0i} =-h_i^2 \omega _i$,
we have
$ds^2 =  g_{\mu \nu} dx^{\mu} dx^{\nu} =-h_0^2 dt^2
  +\sum _{i=1}^3 \left [h_i^2(dx^i)^2 - 2h_i^2 \omega _i dt dx ^i 
\right]$.
When we define the lapse function $\alpha$ and shift vector $\beta^i$ by
$\alpha = \left [ h_0^2+\sum _{i=1}^3 
\left ( h_i \omega _i \right ) ^2 \right ]^{1/2}$,
$\beta ^ {i} = \frac{h_i \omega _i}{\alpha }$, the line element $ds$ is written by
$ds^2=-\alpha ^2 dt^2+\sum _{i=1}^3 (h_i dx^i - \alpha \beta ^i dt)^2$.
We also have $g=-(\alpha h_1 h_2 h_3)^2$.
Using the ``zero-angular-momentum observer (ZAMO) frame" $\hat{x}^\mu$, 
where the line element $ds$ is given by $ds^2 = -d\hat{t}^2 + \sum_i (\hat{x}^i)^2 
= \eta_{\mu\nu} d\hat{x}^\mu d\hat{x}^\nu$, we have the 3+1 formalism 
of the generalized GRMHD and the Maxwell equations.
As for equations including only derivatives of contravariant vectors $A^\mu$
or anti-symmetric 2nd rank tensors $A^{\mu\nu}$, 
we obtain their 3+1 formalism easily
using Equations 
$\nabla_\nu A^\nu = \frac{1}{\sqrt{-g}} \partial (\sqrt{-g} A^\nu)$ or 
$\nabla_\nu A^{\mu\nu} = \frac{1}{\sqrt{-g}} \partial (\sqrt{-g} A^{\mu\nu})$.
%% (\ref{eqcov}) and (\ref{eqast}).
With respect to any equation including a term of derivative of the symmetric
2nd rank tensor,
\begin{equation}
\nabla_\nu S^{\mu\nu} = H^\mu ,
\end{equation}
the 3+1 formalism is given by
\begin{eqnarray}
&& \frac{\partial}{\partial t} \hat{S}^{00}
+ \frac{1}{h_1 h_2 h_3} \sum_j \frac{\partial}{\partial x^j}
\left [ \frac{\alpha h_1 h_2 h_3}{h_j} 
\left ( \hat{S}^{0j} +  \beta^j \hat{S}^{00} \right ) \right ] 
 + \sum_j \frac{1}{h_j} \frac{\partial \alpha}{\partial x^j} \hat{S}^{j0} \nonumber \\
&& + \sum_{j,k}  \alpha \beta^k (G_{kj} \hat{S}^{kj} -G_{jk} \hat{S}^{jj}) 
+ \sum_{j,k} \sigma _{jk} \hat{S}^{jk}   
= \alpha \hat{H}^0 , 
\label{eisaafz}  \\
&& \frac{\partial}{\partial t} \hat{S}^{i0}
+ \frac{1}{h_1 h_2 h_3} \sum_j \frac{\partial}{\partial x^j}
\left [ \frac{\alpha h_1 h_2 h_3}{h_j} 
\left ( \hat{S}^{ij} +  \beta^j \hat{S}^{i0} \right ) \right ] 
 +  \frac{1}{h_i} \frac{\partial \alpha}{\partial x^i} \hat{S}^{00} \nonumber \\
&& -\sum_j \alpha \left [G_{ij} \hat{S}^{ij} - G_{ji} \hat{S}^{jj} 
+ \beta^j (G_{ij} \hat{S}^{0i} -G_{ji} \hat{S}^{0j}) \right ]
+ \sum _j \sigma _{ji} \hat{S}^{0j}   
= \alpha \hat{H}^i ,
\label{eisaaf}
\end{eqnarray}
where $G_{ij} \equiv - \frac{1}{h_i h_j} \frac{\partial h_i}{\partial x^j}$
and $\sigma _{ij} \equiv \frac{1}{h_j} \frac{\partial}{\partial x^j} 
(\alpha \beta^i)$. 

\subsection{Linear analysis of charge separation in stationary disk}

For simplicity, we consider a plasma of a stationary thin disk 
rotating around a Kerr black hole with zero pressure ($p=\Delta p=0$).
The space-time $x^\mu=(t,r,\theta,\phi)$ around the
Kerr black hole with a mass $M$ and rotation parameter $a$ is given by the metrics,
$h_0= (1 - 2 r_{\rm g} r /\Sigma)^{1/2}$, $h_1=\sqrt{\Sigma/\Delta}$, 
$h_2=\sqrt{\Sigma}$, $h_3=\sqrt{A/\Sigma} \sin \theta$,
$\omega_3 = 2 r_{\rm g}^2 a r/A$, and $\omega_i=0$ ($i=1,2$).
Here, $r_{\rm g} = GM$ is the gravitational radius ($G$ is the gravitational
constant), 
$\Delta = r^2 - 2 r_{\rm g} r + (a r_{\rm g})^2$,
$\Sigma = r^2 + (a r_{\rm g})^2 \cos^2 \theta$, and
$A = \{ r^2 + (a r_{\rm g})^2 \}^2 - \Delta (a r_{\rm g})^2 \sin^2 \theta$.
In this metric, the lapse function is $\alpha = \sqrt{\Delta \Sigma/A}$.
The Schwarzschild radius of the black hole is given by $r_{\rm S}=2 r_g$.
The 3-velocity of the circularly rotating disk observed by the ZAMO frame, 
called the Kepler velocity, $V_{\rm K}$, is given by the quadratic equation
\begin{equation}
\alpha G_{31} V_{\rm K}^2 + (\alpha \beta^3 G_{31} + \sigma_{31}) V_{\rm K}
+ \frac{1}{h_1} \frac{\partial \alpha}{\partial r} =0 .
\label{keplervel}
\end{equation}
In investigating linear behavior of charge separation 
in the stationary disk, we assume charge separation is weak, 
$|\rho_{\rm e}'| = |-J^\nu J_\nu| \ll en$. 
Then, we can use an approximation of the enthalpy density
and enthalpy difference density as
\begin{eqnarray}
h^\dagger &\approx& mn , \\
\Delta h^\dagger &\approx& \frac{-m}{2e} \rho_{\rm e}'  , \\
h^\ddagger &\approx& mn + \frac{2\mu m \Delta \mu}{e} \rho_{\rm e}' , \\
\Delta h^\sharp &\approx& mn \Delta \mu + \frac{1-3\mu}{e} m \rho_{\rm e}'  , 
\end{eqnarray}
because of
$n_\pm \approx n \pm \frac{m_\mp}{em} \rho_{\rm e}'$.
The generalized GRMHD equations reduce to
\begin{eqnarray}
&&\nabla_\nu (\rho U^\nu) = 0 , \label{ggrmhdnumlin} \\
&& mn \nabla_\nu  \left [ 
U^\mu U^\nu + \frac{\mu}{(ne)^2} 
\left ( 1 + \frac{2 \mu \Delta \mu \rho_{\rm e}'}{en} \right ) J^\mu J^\nu 
- \frac{\mu \rho_{\rm e}'}{(ne)^2} (U^\mu J^\nu + J^\mu U^\nu ) \right ]
= J^\nu {F^\mu}_\nu ,  \label{ggrmhdmomlin} \\
&& \frac{1}{\omega_{\rm p}^2} \nabla_\nu Q^{\mu \nu}
= \left ( U^\nu - \frac{\Delta \mu}{ne} J^\nu \right) {F^\mu}_\nu 
- \eta \left [ J^\mu - \rho_{\rm e}' (1 + \Theta) U^\mu \right ] , 
\label{ggrmhdohmlin}
\end{eqnarray}
where $\omega_{\rm p} \equiv \sqrt{(ne)^2/(\mu \rho)} = \sqrt{ne^2/(\mu m)}$ 
is the plasma frequency, and $Q^{\mu\nu}$ is approximated by
\begin{equation} 
Q^{\mu \nu} \approx U^\mu J^\nu + J^\mu U^\nu - \rho_{\rm e}' U^\mu U^\nu .
\end{equation}
%% \frac{\mu}{e} 
%% \left ( 1 + \frac{2\mu \Delta \mu \rho_{\rm e}'}{en} \right ) 
%% (U^\mu J^\nu + J^\mu U^\nu ) 
%% - \frac{\mu \rho_{\rm e}'}{e} U^\mu U^\nu
%% - \frac{\mu n}{(ne)^2} 
%% \left \{ \Delta \mu  + \frac{(1-3\mu) \rho_{\rm e}'}{en} \right \} J^\mu J ^\nu .  
%% \]
%% we consider the hydrostatic equilibrium,
%% with respect to the equation of continuity about mass and equation of motion.
%% Here, we assume the electric and magnetic fields are zero in
%% the hydrodynamic equilibrium state.
To perform the linear analysis of the charge separation in the 
hydrostatic equilibrium plasma rotating around the Kerr black hole, 
we consider only the perturbation with respect to the static electric field,
%% to the plasma in the hydrostatic state as,
\begin{eqnarray}
\hat{J}^\mu = \tilde{J}^\mu , && \verb!   ! \hat{\rho}_{\rm e} = \tilde{\rho}_{\rm e} \\
\hat{F}_{i0} = \tilde{E}_i,   && \verb!   ! \hat{F}_{ij} = 0 ,
\end{eqnarray}
where the tildes indicate the infinitesimally small variables,
and we do not consider perturbation to the hydrostatic equilibrium,
\begin{eqnarray}
\rho=\bar{\rho}, \verb!   ! n=\bar{n},  \verb!   !
\hat{U}^\mu = \bar{U}^\mu  .
\end{eqnarray}
In the ZAMO frame, the 4-velocity is given by 
$\bar{U}^0=\gamma_{\rm K}=(1-V_{\rm K}^2)^{-1/2}$, 
$\bar{U}^1=\bar{U}^2=0$, $\hat{U}^3=\gamma_{\rm K} V_{\rm K}$.
The linear analysis requires the Ohm's law (Equation (\ref{onefluidohm})), 
the equation of continuity about current (Equation (\ref{concur})), and 
the Gauss law of electrostatics (temporal component of Equation (\ref{4formamp})),
%% the Maxwell equations with respect to the static electric field among the generalized
%% GRMHD equations (\ref{ggrmhdnumlin})--(\ref{ggrmhdohmlin}) and 
%% the Maxwell equations (\ref{4formfar}) and (\ref{4formamp}),
\begin{eqnarray}
\frac{1}{\omega_{\rm p} ^2}  \nabla_\nu \tilde{Q}^{\mu \nu} &=& 
\bar{U}^\nu {\tilde{F}^\mu}_\nu 
- \eta [\tilde{J}^\mu - (\rho_{\rm e}'+\rho_{\rm e}'\Theta) \bar{U}^\mu],
\label{ohmlawsim} \\
\nabla_\nu \tilde{J}^\nu &=& 0 , 
\label{cntcursim} \\
\nabla_\nu \tilde{F}^{0\nu} &=& \tilde{J}^0 .
\label{amplawsim}
\end{eqnarray}
%
%% The determinant of the matrix with elements $g_{\mu \nu}$ % , $\| g \|$ 
%% is given by
%% $ g \equiv - (\alpha h_1 h_2 h_3)^2$, and
%% the contravariant metric is written explicitly as
%% \begin{equation}
%% g^{00}=- \frac{1}{\alpha ^2}  , \verb!   !
%% g^{i0}=g^{0i}= - \frac{\omega_i}{\alpha^2} =- \frac{\beta^i}{\alpha h_i}, \verb!   !
%% g^{ij} = \frac{1}{h_i h_j} ( \delta ^{ij}
%% -\beta ^i \beta ^j ),
%% \label{defmtc}
%% \end{equation}
%% where $\delta ^{ij}$ is the Kronecker's $\delta$ symbol.
%
Using Equations (\ref{ohmlawsim})--(\ref{amplawsim})
%%  and the formulae (\ref{eqcov}), (\ref{eqast}), 
and (\ref{eisaaf}),
we obtain the 3+1 formalism of the Ohm's law, 
equation of continuity about current, and
Gauss law for the electric field (see also Equations (63) and (67) 
of \citet{koide10}),
\begin{eqnarray}
&& \frac{\partial}{\partial t} \tilde{Q}^{i 0} =
 - \left [  
\frac{1}{h_1 h_2 h_3} \sum_j  \frac{\partial}{\partial x^j} \left ( 
\frac{\alpha h_1 h_2 h_3}{h_j} \left (\tilde{Q}^{ij} + \beta^j \tilde{Q}^{i0}
\right ) \right ) \right . \nonumber \\
&&  \left . +  \frac{1}{h_i} 
\frac{\partial \alpha}{\partial x^i} \tilde{Q}^{00}
- \sum_j \alpha \left \{  G_{ij} \tilde{Q}^{ij} - G_{ji} \tilde{Q}^{jj}
 + \beta^j \left ( G_{ij} \tilde{Q}^{0i} -G_{ji} \tilde{Q}^{j0} \right  ) \right \}
\right ] \nonumber \\
&& + \alpha \omega_{\rm p}^2 \left [ 
\bar{U}^\nu \tilde{F^i}_{\nu}
-\eta [\tilde{J}^i - (\rho_{\rm e}'+\rho_{\rm e}' \Theta) \bar{U}^i] \right ],
\label{genrelgenohm3+1}  
\label{ohm31}  \\
&& \frac{\partial}{\partial t} \tilde{\rho}_{\rm e} = - \frac{1}{h_1 h_2 h_3}
\sum_j \frac{\partial}{\partial x^j} \left (\frac{h_1 h_2 h_3}{h_j} \tilde{J}^j \right ) , \\
&&\tilde{\rho}_{\rm e} = \sum _{j} \frac{1}{h_1 h_2 h_3}
\frac{\partial}{\partial x^j} \left (
\frac{h_1 h_2 h_3}{h_j} \tilde{E}_j
\right )    .
\label{dive}
\label{gle31}
\end{eqnarray}
Here, we used the following approximation,
\begin{eqnarray}
\tilde{Q}^{00} &\approx& \gamma_{\rm K} 
( 2 \tilde{\rho}_{\rm e} - \gamma_{\rm K} \tilde{\rho}_{\rm e}'), \\
\tilde{Q}^{i0} &=& \tilde{Q}^{0i} \approx \gamma_{\rm K} \tilde{J}^i 
+ (\tilde{\rho}_{\rm e} - \gamma_{\rm K} \tilde{\rho}_{\rm e}' ) \bar{U}^i,
\label{mcude} \\
\tilde{Q}^{ij} &\approx& U_{\rm K} [ \delta^{3i} \tilde{J}^j +  \delta^{3j} \tilde{J}^i 
- U_{\rm K} \delta^{i3} \delta^{j3} (\gamma_{\rm K} \tilde{\rho}_{\rm e}
- U_{\rm K} \tilde{J}^3) ],
\label{mcdde}
\end{eqnarray}
where $\tilde{Q}^{00}$ can be regarded as modified charge density and 
$\tilde{Q}^{i0}$ corresponds to the modified current density.
When we use the relation
\[
\rho_{\rm e}' = J^\nu U_\nu = \gamma \rho_{\rm e} - \VEC{J} \cdot \VEC{U}
\approx \gamma_{\rm K} \tilde{\rho}_{\rm e} - \bar{\VEC{U}} \cdot \tilde{\VEC{J}},
\]
where $\VEC{J} \equiv (J^1,J^2,J^3)$ and $\VEC{U} \equiv (U^1,U^2,U^3)$
are the 3-current density and 3-velocity, respectively, we have
\begin{eqnarray}
\hat{Q}^{00} &\approx& \gamma_{\rm K} [ (1-U_{\rm K}^2) \tilde{\rho}_{\rm e}
+ \gamma_{\rm K} U_{\rm K} \tilde{J}^3]   ,\\
\hat{Q}^{i0} &\approx& \gamma_{\rm K} [\tilde{J}^i + U_{\rm K}^2 \delta^{i3} \tilde{J}^3
-U_{\rm K}^3 \delta^{i3} \tilde{\rho}_{\rm e}] .
\end{eqnarray}
%
%% When we consider the space-time around the Schwarzschild black hole,
%% $x^\mu =(t,r,\theta,\phi)$, the metric are given by
%% \begin{equation}
%% h_0 = \alpha, \verb!   ! h_1 = \frac{1}{\alpha}, \verb!   !, h_2=r, 
%% \verb!   ! h_3 = r \sin \theta,
%% \end{equation}
%% where $\alpha = (1-r/r_{\rm S})^{1/2}$. 
%% We have $\hat{U}^i = U_{\rm K} \delta^{i3}$, $U_{\rm K} = V_{\rm K}/\sqrt{1-V_{\rm K}^2}
%% =1/\sqrt{2r-3}$.
Because in the present linear analysis, we can assume the quasi-charge neutrality,
thus we have an approximation of $\rho_{\rm e}' \Theta$ as,
\begin{equation}
\rho_{\rm e}' \Theta \approx \frac{\Delta \mu}{2ne} |\VEC{J}'|^2,
\end{equation}
where $|\VEC{J}'|^2 = (\gamma - 1) \rho_{\rm e}^2
- 2 \gamma \rho_{\rm e} \VEC{U} \cdot \VEC{J} + |\VEC{J}|^2
+ (\VEC{U} \cdot \VEC{J})^2$. \citep{koide10}
Then, when $J'$ is infinitesimally small, we have $\rho_{\rm e}' \Theta \approx 0$.

Here, we assume the perturbation is symmetric with respect to the polar axis 
and the equatorial plane,
$\partial/\partial x^i = \delta^{1i} \partial/\partial x^1$.
Then, the equations with respect to the perturbation of the charge separation at
the equatorial plane are as follows:
\begin{eqnarray}
\frac{\partial}{\partial t} \tilde{J}^1 &=& 
[\alpha G_{31} ( 2 V_{\rm K} + \beta^3) + \sigma_{31}] 
(V_{\rm K} \tilde{\rho}_{\rm e} - \tilde{J}^3)
+ \frac{\alpha \omega_{\rm p}^2}{\gamma_{\rm K}} 
(\gamma_{\rm K} \tilde{E}_1 - \eta \tilde{J}^1) , 
\label{lineqj1} \\
\frac{\partial}{\partial t} (\tilde{J}^3 - \hat{V}_{\rm K}^3 \tilde{\rho}_{\rm e})
&=& - \frac{1}{h_1 h_2 h_3^2 \gamma_{\rm K}^3} \frac{\partial}{\partial r}
(\alpha h_2 h_3^2 \hat{U}_{\rm K} \tilde{J}^1 )   \\
&& + \omega_{\rm p}^2 \frac{\alpha}{\gamma_{\rm K}^2} 
\left [ \tilde{E}_3 
- \eta \left (\gamma_{\rm K} \tilde{J}^3
- U_{\rm K} \tilde{\rho}_{\rm e} \right ) \right ] , \\
\frac{\partial \tilde{\rho}_{\rm e}}{\partial t} &=& 
-\frac{1}{h_1 h_2 h_3} \frac{\partial}{\partial r} (\alpha h_2 h_3 \tilde{J}^1) ,  \\
\tilde{\rho}_{\rm e} &=& 
\frac{1}{h_1 h_2 h_3} \frac{\partial}{\partial r} (h_2 h_3 \tilde{E}^1) .
\end{eqnarray}
To derive Equation (\ref{lineqj1}), we used Equation (\ref{keplervel}).
For simplicity, we assume the wave length of the perturbation is much smaller
than the characteristic scale-length of the metrics and
the Keplerian rotation around the
black hole, $\sim r_{\rm S}$, and we put the perturbation is proportional to 
$\exp (ikr - i \omega t)$.
Here, we note that we have to consider the relation between the derivatives
\begin{equation}
\frac{1}{\alpha U_{\rm K} h_1 h_2 h_3^2}
\frac{\partial}{\partial r} (\alpha h_2 h_3^2 U_{\rm K} \tilde{J}^1) 
- \frac{1}{\alpha h_1 h_2 h_3}
\frac{\partial}{\partial r} (\alpha h_2 h_3^2 \tilde{J}^1) 
%% = \frac{\tilde{J}^2}{h_3 U_{\rm K}} \frac{\partial}{\partial r} (h_3 U_{\rm K})
= \frac{\tilde{J}^1}{L_{\rm K}} \frac{\partial L_{\rm K}}{\partial r} ,
\end{equation}
where $L_{\rm K} = h_3 U_{\rm K}$ is the specific angular momentum of the Keplerian disk.
Finally, we obtain the dispersion relation of the charge separation in the plasma
disk rotating circularly around the Kerr black hole,
\begin{equation}
\left ( \frac{\omega}{\alpha} \right )^3 + 2 i \eta' \left ( \frac{\omega}{\alpha} \right )^2
- \left [ \eta'^2 + \omega_{\rm p}^2 - 2 g_{\rm K} \Lambda_{\rm K} \right ] 
 \frac{\omega}{\alpha} 
- i \eta' \omega_{\rm p}^2 = 0 ,
\label{findisrel0}
\end{equation}
where $\eta' = \omega_{\rm p}^2 \eta/\gamma_{\rm K}$,
\begin{eqnarray}
g_{\rm K} &=& \frac{V_{\rm K}}{2 \alpha} \left [ 
\alpha G_{31} ( 2 V_{\rm K} + \beta^3 ) + \sigma_{31} \right ] 
  = - \frac{1}{h_1} \frac{\partial}{\partial r} \log \alpha
- \frac{V_{\rm K}}{2 \alpha} (\alpha G_{31} \beta^3 + \sigma_{31}) , \\
\Lambda_{\rm K} &=& \frac{1}{h_1 \gamma_{\rm K}^2 L_{\rm K}} 
\frac{\partial L_{\rm K}}{\partial r}   .
\end{eqnarray}
Here, it is noted that the stability condition of the accretion disk is given 
by $\partial L_{\rm K}/\partial r > 0$ ($\Lambda_{\rm K} > 0$),
and the condition $\partial L_{\rm K}/\partial r = 0$ ($\Lambda_{\rm K} = 0$) yields 
the radial coordinate of the ISCO, $r=r_{\rm ISCO}$.
%% where $\omega_{\rm p} = \sqrt{e^2 n /(\mu m)}$ is the plasma frequency,
%% $L_{\rm K}=h_3 \hat{U}_{\rm K}=r/\sqrt{r_{\rm S}(2r-3r_{\rm S})}$ is
%% the specific angular momentum of Kepler disk,
%% and $\omega_{\rm r}=\alpha \omega_{\rm p}^2 \eta /\gamma_{\rm K}$ is the variable
%% proportional to the resistivity.
%% When we use the frequency observed by the ZAMO frame, $\hat{\omega} = \omega/\alpha$ 
%% and $\hat{\omega}_{\rm r} = \omega_{\rm r}/\alpha$, we obtain the cubic equation of
%% dispersion relation of the charge separation mode:
%% \begin{equation}
%% \hat{\omega}^3 + i \hat{\omega}_{\rm r} \{ 1 + (1+\Theta) \hat{V}_{\rm K}^2 \} \hat{\omega}^2
%% - \{ \hat{\omega}_{\rm r}^2 (1+\Theta)\hat{V}_{\rm K}^2 + \omega_{\rm p}^2 + 2 g \Lambda \} 
%% \hat{\omega}
%% + 2 g k \hat{\omega}_{\rm r} \Theta - i \omega_{\rm p}^2 \hat{\omega}_{\rm r}
%% (1+\Theta) \hat{V}_{\rm K}^2 = 0,
%% \label{disrelcsi}
%% \end{equation}
%% where
%% \[
%% \Lambda = \frac{1}{h_1} \frac{\partial \ln L_{\rm K}}{\partial r} \verb!   ! {\rm and}
%% \verb!  ! g= \frac{1}{\gamma^2} \frac{1}{h_1} \frac{\partial \ln \alpha}{\partial r}.
%% \]

In the case of zero resistivity ($\eta=0$), Equation (\ref{findisrel0}) yields the dispersion
relation of the charge separation in the plasma disk as
\begin{equation}
\omega^2 = \alpha^2 ( \omega_{\rm p}^2 - 2 g_{\rm K} \Lambda_{\rm K} ) 
\equiv \omega_0^2  .
\label{findisrel1}
\end{equation}
%% The radius of the ISCO around the Schwarzschild black hole is $3r_{\rm S}$, 
%% and then, the condition of unstable region of the disk is given by 
%% $\partial L_{\rm K}/\partial r < 0$. 
In the region inside of the ISCO, when 
\begin{equation}
\omega_{\rm p} < \sqrt{2g_{\rm K} \Lambda_{\rm K}} ,
%% = \sqrt{-\frac{2}{\gamma_{\rm K}^2} \frac{1}{h_1} 
%% \frac{\partial \ln \alpha}{\partial r}
%% \frac{\partial \ln L_{\rm K}}{\partial r}},
\label{cricsiide}
\end{equation}
the charge separation becomes unstable.
However, it is noted that the disk is unstable with the growth rate
$\gamma_{\rm disk} = \alpha \sqrt{2 g_{\rm K} \Lambda_{\rm K}}$, which is larger than
the growth rate of the charge separation instability. In the very low plasma
density case ($\omega_{\rm p} \ll \gamma_{\rm disk}$), the growth rate
of the disk instability and the charge separation instability become comparable.
Then, in this situation, the charge separation instability may appear in the
disk falling into the black hole and may make the black hole charged.
On the other hand, in the other usual plasma density case, 
the charge separation instability
is forbidden or inhibited behind the disk instability.

To investigate the effect of resistivity, we consider a solution for a very weak
resistivity limit, $\alpha \eta' \ll \omega_0$. In this approach, 
we treat the difference $\Delta \omega = \omega - \omega_0$
is an infinitesimal variable which is comparable to $(\alpha \eta'/\omega_0^2) \omega_0$.
The dispersion relation (\ref{findisrel0}) yields
\begin{equation}
2 \omega_0^2 \Delta \omega
+ i \eta' \left [ 2 \omega_0^2 -  \alpha^2 \omega_{\rm p}^2
\right ] = 0,
\end{equation}
and we have the solution
\begin{equation}
\omega = \omega_0 - i \frac{\alpha^2 \eta'}{2\omega_0^2}
\left ( \omega_{\rm p}^2 - 4 g_{\rm K} \Lambda_{\rm K} \right ) 
=\alpha ( \omega_{\rm p}^2 - 2 g_{\rm K} \Lambda_{\rm K})^{1/2}
-i \frac{\alpha^2 \eta' (\omega_{\rm p}^2 
- 4 g_{\rm K} \Lambda_{\rm K})}{\omega_{\rm p}^2 - 2 g_{\rm K} \Lambda_{\rm K}} .
\label{cricsires}
\end{equation}
It is noted that even in the case of $\omega_0^2 > 0$ 
(stable state in the zero resistivity case),
$\omega_{\rm p}^2 -4 g_{\rm K} \Lambda_{\rm K}$ can become negative
in the unstable region ($\Lambda_{\rm K} < 0$), which means the resistivity induces
the charge separation instability, while in the outside of the ISCO, $r > r_{\rm ISCO}$, 
the instability is forbidden.
% This clearly shows that the resistivity reduces the condition of the charge separation 
% instability.
On the other hand, in the case of $\omega_0^2 < 0$, the resistivity stabilizes
the charge separation instability as shown by the last term of the right-hand
side of Equation (\ref{cricsires}).

\section{Discussion \label{sec3}}

We have shown the charge separation instability of the circularly rotating
plasma inside the ISCO ($r \le r_{\rm ISCO}$) around the Kerr black hole.
This instability is forbidden in the stable disk outside of the ISCO.
Furthermore, even in the unstable disk region inside of the ISCO, the growth rate
of the charge separation instability
is smaller than that of the disk instability. However, when the plasma density
is much lower than the critical density, growth rates of
the charge separation and disk instabilities become comparable and the charge 
separation instability becomes apparent. Then, the charge separation 
instability makes
the disk plasma falling into the black hole charged and may 
eventually charge the hole.
%% The instability happens in the low density plasma whose density $n$ 
The critical plasma density is expressed as
\begin{equation}
n_{\rm crit} \equiv 
\frac{2 \mu m}{e^2}  g_{\rm K} \Lambda_{\rm K}  ,
\label{conparnum}
\end{equation}
where the critical density $n_{\rm crit}$ is given by Equation (\ref{cricsiide}) 
with $\omega_{\rm p}^2 = n_{\rm crit} e^2/(\mu m)$ in the zero resistivity case.
With respect to the resistive case, Equation (\ref{cricsires}) indicates that 
the weak resistivity makes the charge separation instability happen easier in
the stable region of the charge separation instability inside of the ISCO,
while in the unstable region of the ideal MHD situation, the resistivity stabilizes 
the instability.
In a case with resistivity, the growth rate of the charge separation instability
is also smaller than that of the disk instability inside of the ISCO.
Then, unless the plasma density $n$ is much less than the critical density $n_{\rm crit}$
so that the growth rate of the charge separation instability is comparable to that
of the disk instability, the charge separation instability is hidden behind the disk
instability inside of the ISCO and is also forbidden in the stable disk around 
the astrophysical black holes.
However, we emphasize again that when the plasma density is much less than 
the critical density, $n \ll n_{\rm crit}$,
the charge separation instability may be apparent as the charged disk falling into the black
hole. 
%% The fall of the charged disk will produce a charged black hole (Kerr-Newman black hole),
%% which suggests the spontaneous formation of the Kerr-Newman black hole.

\begin{figure}
\includegraphics{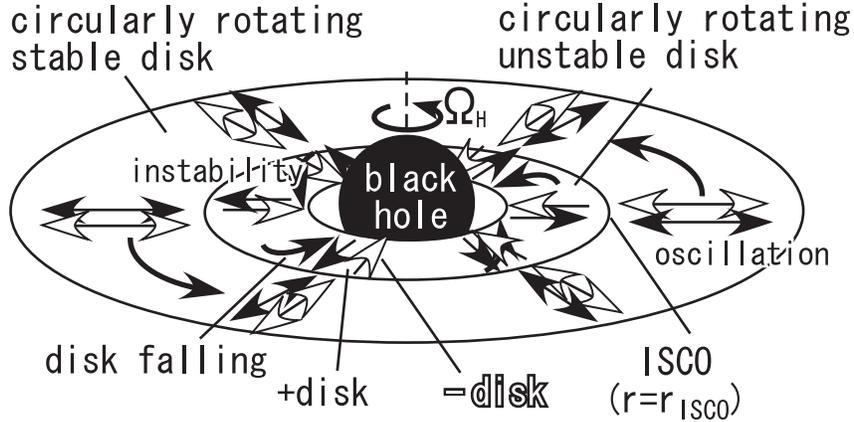}% Here is how to import EPS art
\caption{
Schematic picture of the charge separation instability of the plasma
in an unstable region around a black hole ($r<r_{\rm ISCO}$). 
The neutral disk consists of positively/negatively 
charged disks (+/-- disks); when the electric field due to the charge separation is 
not strong enough to suppress the charge separation, 
for example, the positive disk shifts outward and the negative disk falls into
the black hole exponentially in the unstable disk region.
The reverse is also true. This shift and falling cause the
exponential charge separation.
%% instability occurs in the unstable region of the $\pm$ disks.
%% In the region, the disk is unstable and falls into the black hole with the growth
%% rate larger than that of the charge separation instability. Then, in ordinal astrophysical
%% situations, the charge separation instability is hidden behind the disk instability.
%% However, when the plasma density of the disk is much lower than the critical density,
%% the both growth rates become comparable and the charge separation instability becomes
%% significant.
\label{fig1}}
\end{figure}

The mechanism of the charge separation instability is 
explained by the following schematic picture.
Consider that the accretion disk is composed of two disks of positively charged particles
(+ disk) and of negatively charged particles (-- disk) (Fig. \ref{fig1}).
Both of the purely charged disks ($\pm$ disks) are unstable in the region $r<r_{\rm ISCO}$.
However, even in the unstable region, when one disk ($\pm$ disk) falls and 
the other disk ($\mp$ disk) shifts outward,
the electric field is induced and tends to suppress the charge separation.
When the electric field is strong enough, it causes plasma oscillation.
On the other hand, when the electric field can not become so strong 
(in the case of $n \ll n_{\rm crit}$),
the suppression is not strong (effective) enough and the charged disks separate each other 
increasingly so that the charge separation is induced exponentially.
%% This is just the charge separation instability newly shown in this paper.
In the outer region of the ISCO, $r>r_{\rm ISCO}$, 
the plasma oscillation is always induced by the charge separation because of 
the stability of the $\pm$ disks.
Using this intuitive picture of the charge separation instability, 
we suggest that 
the charge separation instability with small wave number $k$ 
%% (global part of the disk) 
is also possible in the unstable disk around the rotating black hole, 
when the plasma oscillation frequency $\omega_{\rm p}$
is small enough and the plasma density is sufficiently small.
This picture also suggests that in a magnetized plasma disk, 
the similar instability
of charge separation would be caused to break the free charge quasi-neutrality
when the purely charged disk is gravitationally unstable and the plasma
density is low enough.
%% which is required for the charge separation instability.
This charge separation grows exponentially until the nonlinear effects
begin to suppress the instability. For example, the disk charged
by the charge separation instability falls into the black hole
and charges the black hole. The electric field of the charged black hole
will suppress the fall of the charged disk and  induce the fall of 
the oppositely charged plasma around the black hole.

The charge separation instability may also happen in the pulsar magnetosphere 
when the radius of the ISCO is larger than the central star radius.
However, when the magnetic field is extremely strong,
the free charge due to the charge separation instability 
would not be induced because 
the magnetic field suppresses the gravitational instability of the disk.
Plasma dynamics of the pulsar magnetosphere outside of the relativistic
star were investigated with the Schwarzschild metric 
by Henriksen and Rayburn \cite{henriksen74}.
They discussed the net charge separation and the instability, while
they did not perform the linear analysis of the charge separation 
in bulk plasma.
They considered the large-scale net charge separation within the free
charge quasi-neutrality and the charge separation instability 
produced by the current driven plasma turbulence which
leads charge fluctuation on small scales less than the Debye length.

Here, we estimate the critical density of the charge separation instability 
(Equation (\ref{conparnum})) for an individual astrophysical object. 
For a rough estimation of the instability condition, we consider
the cases of the Schwarzschild black holes with the mass $M_{\rm BH}$ ($a=0$).
The critical density of the instability is expressed
in the MKSA system of units (SI unit) by
\begin{equation}
n_{\rm crit} = \frac{\mu m r_{\rm S}}{2 \mu_0 e^2} 
\frac{3r_{\rm S} -r}{\sqrt{r^5 (r-r_{\rm S})^3}}  ,
\end{equation}
where $\mu_0$ is the magnetic permeability in vacuum.
%%  and $r_{\rm S} \equiv 2 r_{\rm g}$ is the Schwarzschild radius. 
We estimate it at $r=2r_{\rm S}$ to get
\begin{equation}
n_{\rm crit}^* = \frac{\mu m}{m_{\rm e}}
\frac{m_{\rm e}}{2^{7/2} \mu_0 e^2 r_{\rm S}^2}
= 2.5 \times 10^{12}
\left ( r_{\rm S} [{\rm m}] \right ) ^{-2}
\frac{\mu m}{m_{\rm e}} \,\,\,  [{\rm m}^{-3}]  
= 2.8 \times 10^5 \left ( \frac{M_{\rm BH}}{M_\odot} \right )^{-2}
\frac{\mu m}{m_{\rm e}},
\label{ncrit}
\end{equation}
where $m_{\rm e}$ is the electron mass. If we assume the electron-proton
and electron-positron plasmas, we have $\mu m/m_{\rm e}=1$ and $\mu m/m_{\rm e}=1/2$, 
respectively. In this paragraph, we set $\mu m/m_{\rm e}=1$.
We use values of the typical density $n$
of accretion disks around black holes of individual objects listed up by 
\citet{koide10}.
When we consider the active galactic nucleus (AGN) of M87 
whose central black hole mass
is $M_{\rm BH} = 3 \times 10^9 M_\odot$
($r_{\rm S}=9 \times 10^{12} {\rm m}$) \citep{macchetto97},
the critical density of the charge separation instability is estimated as
$n_{\rm crit}^* = 3 \times 10^{-14} \, [{\rm m}^{-3}]$. This value is extremely small
compared not only to the estimated value at the accretion disk around the black hole of
$n = 8 \times 10^{21} \, [{\rm m}^{-3}]$, but also to
the averaged particle number density in the extragalactic region, 
$n_{\rm U} \sim 10 \, [{\rm m}^{-3}]$. \citep{weinberg72}
In the case of Sgr A$^*$, a supermassive black hole in Galaxy, whose central black hole
mass is $M_{\rm BH} = 4.4 \times 10^6 M_\odot$ 
($r_{\rm S}=1.3 \times 10^{10} {\rm m}$) \citep{genzel10},
the critical density is $n_{\rm crit}^* = 1.4 \times 10^{-8} \, [{\rm m}^{-3}]$. 
This is also extremely small compared to the value at the accretion disk,
$n = 9 \times 10^{23} \, [{\rm m}^{-3}]$. As we estimated above, the charge separation instability
hardly happens around the supermassive black holes.
In the case of black hole X-ray binaries, for example, the micro-quasar, GRS1915+105, 
whose central black hole mass is
$M_{\rm BH} = 14 M_\odot$ ($r_{\rm S}=4.2 \times 10^{4} {\rm m}$) \citep{greiner01},
the critical density is $n_{\rm crit}^* = 1.4 \times 10^{3} \, [{\rm m}^{-3}]$.
This density is much smaller than the density at the accretion disk around the
black hole, $n = 4 \times 10^{27} \, [{\rm m}^{-3}]$. However, 
a density smaller than
the critical density may be realized in a low density disk around 
a single stellar-mass black hole, for example. In the case of a very low
density disk, the charge separation instability may be caused, while it would
be difficult to observe because of its low activity of the thin disk.

The charge instability causes the falling of the charged plasma
into the black hole and will charge the black hole.
Here, we estimate the influence of the electric field of the charged
black hole to the ambient plasma in the following case.
We assume that the one-component charged fluid with the height 
$H \sim r_{\rm S}$,
the inner radius $r_{\rm inner} \sim 2 r_{\rm S}$, the
outer radius  $r_{\rm outer} \sim 2 r_{\rm S} + L$, and
the density $n \sim n_{\rm crit}^*$ falls 
due to the charge separation instability
and is swallowed by the black hole.
The charge contained in the one-component fluid
is $Q \sim 8 \pi r_{\rm S}^2 L n_{\rm crit}^* e$.
The electric and gravitational forces which act on a charged
particle with the mass $m_-$ and the charge $-e$ located at
$r=r$ are $F_{\rm E} \sim \frac{1}{4 \pi \epsilon_0} \frac{e Q}{r^2}$
and $F_{\rm grav} \sim G \frac{m M_{\rm BH}}{r^2} $, respectively.
We estimate the disk scale $L$ where the electric and gravitational
forces become comparable: $F_{\rm E} \sim F_{\rm grav}$.
It yields $L \sim \frac{\epsilon_0 m_- c^2}{4 e^2 n_{\rm crit}^* r_{\rm S}}$.
Using Equation (\ref{ncrit}), we have the simple expression,
\begin{equation}
\frac{L}{r_{\rm S}} \sim 2^{3/2} \frac{m}{m_+} \sim 2.8 ,
\end{equation}
where the ratio does not depend on the black hole mass, $M_{\rm BH}$.
This suggests that the small disk with the scale of the Schwarzschild
radius supplies so much charge to the black hole through
the charge separation instability within short time scale of the instability
that the charged black hole
influences the plasma dynamics around it.
This electric field will suppress the further charging of the black hole.
%%  due to the charge separation instability.

It is true that the charge separation instability is caused only 
in the unstable disk region. 
However, the current induced by the instability may reach 
the stable disk region because of the inertia of the current. 
The current supplies the net charge in the 
stable disk around the black hole.
Thus, the charge separation due to the instability
may induce a distinctive drastic phenomena in the plasma of the stable disk
regions around black holes, while the net charge may be canceled by the charge
supply from the outer disk.
For example, the gravitational magnetic reconnection can be induced by the charge
separation \citep{koide10}. 
If the charge separation is significant, it also causes a strong electric field. 
%% It also suggests that we have to treat the plasma inside of the ISCO as the strongly 
%% charged plasma and the charge separation will induce the strong electric field. 
This strong electric field may accelerate 
particles which can explain high energy cosmic rays as well as in the ``outer gap"
of the pulsar magnetosphere \citep{holloway73,holloway81}.
The cause of these distinctive phenomena of plasma around black holes will be clarified with
more detailed analysis of the generalized GRMHD equations and numerical simulations.
The numerical simulations of the generalized GRMHD beyond the ideal GRMHD \citep[e.g.,][]{koide06,mckinney06}
and the resistive relativistic MHD with the acausal relativistic Ohm's law
\citep{watanabe06} would provide a useful and essential tool for such analysis.
This is our next subject.

In the last paragraph of this paper, we suggest a radiation process associated with
the charge separation instability. The charge separation instability does not
depend on the wave number $k$ in the zero resistivity case as shown by Equation (\ref{findisrel0}).
Then, very strong electric field with a very large wave number can be caused 
through the instability. When a high energy charged particle passes through the
strong electric field, the particle is accelerated and decelerated reciprocally, and
emits strong radiation. This radiation should be detected around the inner edge of
disks of black holes.

%% \begin{figure}
%% \epsscale{1.0}
%% \plotone{fig1.eps}
%% \caption{
%% Schematic picture of the charge separation instability of the plasma
%% in an unstable region around a black hole ($r<r_{\rm ISCO}$). 
%% The neutral disk consists of positively/negatively 
%% charged disks (+/-- disks); when the electric field due to the charge separation is 
%% not strong enough to suppress the charge separation, 
%% for example, the positive disk shifts outward and the negative disk falls into
%% the black hole exponentially in the unstable disk region.
%% The reverse is also true. This shift and falling cause the
%% exponential charge separation.
%% \label{fig1}}
%% \end{figure}

\begin{acknowledgments}
I am grateful to Mika Koide for her helpful comments on this paper.
I thank Kunihito Ioka for the fruitful discussion, which inspired me to execute
this analysis of the charge separation instability around black holes.
This work was supported in part by the Science Research Fund of
the Japanese Ministry of Education, Culture, Sports, Science and Technology.
\end{acknowledgments}

%% \newpage
\appendix
\section{Basis of generalized GRMHD equations
\label{appenda}}

The generalized GRMHD equations are derived from the general relativistic
two-fluid equations of plasma, which is composed of positively charged particles
with charge $e$ and mass $m_+$ and negatively charged particles with charge $e$
and mass $m_-$ \citep{koide10}. 
The variables of the generalized GRMHD equations are defined
by the average and difference of the variables of the two-fluid equations.
%% As \citet{koide10}, 
We note a variable of fluid composed by the positively
charged particles with a subscript ``+" and that of negative fluid with ``--".
We write the variables of two fluids as: $n_\pm$ is the number density,
$\gamma_\pm'$ is the Lorentz factor observed by the center of mass frame 
of the charged fluids, $p_\pm$ is pressure, $h_\pm$ is the enthalpy density,
$U_\pm^\mu$ is 4-velocity. 
%% The typical mass of the plasma particle is given by $m=m_+ + m_-$. 
The variables of the generalized GRMHD equations are based as follows:
\begin{eqnarray}
\rho &=& m_+ n_+ \gamma_+' + m_- n_- \gamma_-'  ,  \\
n &=& \frac{\rho}{m},  \\
p &=& p_+ + p_- ,  \\
\Delta p &=& p_+ - p_- ,  \\
U^\mu &=& \frac{1}{\rho} (m_+ n_+ U_+^\mu + m_- n_- U_-^\mu) ,  \\
J^\mu &=& e( n_+ U_+^\mu -  n_- U_-^\mu)  ,  \\
h^\dagger &=& n^2 \left ( \frac{h_+}{n_+^2} + \frac{h_-}{n_-^2} \right ) , \\
\Delta h^\dagger &=& \frac{m n^2}{2} \left ( \frac{h_+}{m_+ n_+^2} 
- \frac{h_-}{m_- n_-^2} \right ) . 
\end{eqnarray}
%%
%% In addition of these basic variables, 
We also use the variables 
with respect to the enthalpy density:
\begin{eqnarray}
h^\ddagger &=& \frac{n^2}{4 \mu} \left [  
\frac{h_+}{n_+^2} \left ( \frac{2m_-}{m} \right )^2
+ \frac{h_-}{n_-^2} \left ( \frac{2m_+}{m} \right )^2
\right ]  = h^\dagger - \Delta \mu \Delta h^\dagger  , \\
\Delta h^\sharp &=&  - \frac{n^2}{8 \mu} \left [ 
\frac{h_+}{n_+^2} \left ( \frac{2m_-}{m} \right )^3
- \frac{h_-}{n_-^2} \left ( \frac{2m_+}{m} \right )^3
\right ]  
=\Delta \mu h^\dagger - \frac{1-3\mu}{2 \mu} \Delta h^\dagger  .
\end{eqnarray}
Incidentally, we sometimes assume the plasma consists of two perfect fluids 
with the equal specific heat ratio, $\Gamma$. The equations of states are
\begin{eqnarray}
h^\dagger &=& n^2 \left [ \frac{m_+}{n_+} + \frac{m_-}{n_-}  +
\frac{\Gamma}{2 (\Gamma -1)} \left \{ 
\left ( \frac{1}{n_+^2} + \frac{1}{n_-^2} \right ) p 
+ \left (  \frac{1}{n_+^2} -  \frac{1}{n_-^2}\right ) \Delta p
\right \} \right ] , \\
\Delta h^\dagger &=& 2 \mu m n^2 \left [ \frac{1}{n_+} - \frac{1}{n_-}  
\right .
\nonumber \\
&& \left . + \frac{\Gamma}{2 (\Gamma -1)} \left \{ 
\left ( \frac{1}{m_+ n_+^2} - \frac{1}{m_- n_-^2} \right ) p 
+ \left (  \frac{1}{m_+ n_+^2} +  \frac{1}{m_- n_-^2}\right ) \Delta p
\right \} \right ] ,
\label{eosdhi}
\end{eqnarray}
where
\begin{equation}
n_\pm \equiv \left [ n^2 \mp \frac{2 m_\mp n}{em} U^\nu J_\nu 
-\left ( \frac{m_\mp}{em} \right )^2 J^\nu J_\nu \right ]^{1/2}  ,
\label{defnpm}
\end{equation}
corresponds to the particle number density of each charged fluid
(see Equations (74) -- (78) of \citet{koide10})\footnote{Equations (76) 
of \citet{koide10} contain several typographical errors. Here, we correct them
in Equation (\ref{defnpm}). The variable $\rho_\pm$ in Equation (76) 
of \citet{koide10} is given
by $\rho_\pm = m n_\pm$.}.

The energy-momentum tensors $T^{\mu\nu}$ and the charge-current density
tensor $Q^{\mu\nu}$ are given by
\begin{eqnarray}
T^{\mu\nu} &=& T^{\mu\nu}_+ + T^{\mu\nu}_- + T^{\mu\nu}_{\rm EM} , \\
Q^{\mu\nu} &=& \frac{en}{h^\dagger} 
\left ( \frac{1}{m_+} T^{\mu\nu}_+ - \frac{1}{m_-} T^{\mu\nu}_- \right ) , 
\end{eqnarray}
where $T^{\mu\nu}_\pm = g^{\mu\nu} p_\pm + h_\pm U^\mu_\pm U^\nu_\pm$
are the energy-momentum tensor of the two fluids and 
$T^{\mu\nu}_{\rm EM} = {F^\mu}_\sigma F^{\nu\sigma} 
- \frac{1}{4} g^{\mu\nu}  F^{\kappa\lambda} F_{\kappa\lambda}$
is the Maxwell stress tensor.

The relativistic two-fluid equations come from the continuity equations
of particle number and conservation law of energy and momentum:
\begin{eqnarray}
\nabla_\nu (n_\pm U_\pm^\nu ) &=& 0, \\
\nabla_\nu (h_\pm U_\pm^\mu U_\pm^\nu ) &=& - \nabla^\mu p_\pm
\pm e n_\pm {U_\pm^\mu}_\nu \pm R^\mu,
\end{eqnarray}
where $R^\mu$ is the frictional 4-force density between the two fluids.
It connected with the resistivity and current as
\begin{equation}
R^\mu = - \eta n e [J^\mu - \rho_{\bf e}' ( 1+ \Theta) U^\mu].
\end{equation}


\end{document}